\documentstyle[multicol,prl,aps,epsf]{revtex}
\begin{document}
\tightenlines
\title{Quantum Fluctuations and Dissipation in Thin Superconducting Wires}
\author{Andrei D. Zaikin$^{1,2}$, Dmitrii S. Golubev$^{2,3}$,
        Anne van Otterlo$^{4}$, and Gergely T. Zim\'{a}nyi$^{4}$}
\address{$^1$ Institut f\"{u}r Theoretische Festk\"orperphysik,
Universit\"at Karlsruhe, 76128 Karlsruhe, FRG\\
$^2$ I.E.Tamm Department of Theoretical Physics, P.N.Lebedev
Physics Institute, Leninskii pr. 53, 117924 Moscow, Russia\\
$^{3}$ Physics Department, Chalmers University of Technology, 
S-41296 G\"oteborg, Sweden\\
$^4$ Physics Department, University of California, Davis,
CA 95616, USA}

\maketitle

\begin{abstract}
We investigate quantum fluctuations in thin superconducting wires. We
demonstrate that quantum phase slips dominate the system behavior 
at low temperatures and are well in the measurable range for sufficiently
thin wires. We discuss the effect of dissipation, predict a new quantum
superconductor-to-metal (insulator) phase transitions for wires with
thicknesses in the 10-nm range, evaluate the resistance $R(T)$ of 
such wires and compare our results with recent experimental findings.
\end{abstract}


\begin{multicols}{2}

It is well known that fluctuations wash out the long-range order in
low dimensional superconductors \cite{HMW}. Does this result mean that
the resistance of such superconductors always remains finite (or even
infinite), or can it drop to zero under certain conditions?
A lot is known about the behavior of two-dimensional (2D) superconducting
films where physics is essentially determined by the
Kosterlitz-Thouless-Berezinskii (KTB) phase transition \cite{KTB}.
In quasi-1D superconducting wires below the (mean field) critical
temperature $T_C$ a nonzero resistivity can be caused by thermally activated
phase slips (TAPS) \cite{LAMH}. This effect is of practical importance
at temperatures close to $T_C$ where the theoretical predictions have been
verified experimentally \cite{Exp}.
However, as the temperature is lowered the number of TAPS decreases
exponentially and no measurable resistance is predicted by the
theory~\cite{LAMH} at $T$ not very close to $T_C$.
Nevertheless, the experiments by Giordano~\cite{Gio} clearly demonstrate
a notable resistivity of ultra-thin superconducting wires far below $T_{C}$.
More recently strong deviations from the TAPS
prediction in thin (quasi-)1D wires have been also demonstrated in
other experiments~\cite{Dynes}.

The natural explanation of these observations is in terms of quantum
fluctuations which generate quantum phase slips (QPS) in 1D superconducting
wires. However, first estimates for the QPS tunneling rate derived
from the time-dependent Ginzburg-Landau based theories \cite{sm,Duan}
turned out to be by far too small to explain the experimental findings
\cite{Gio} (see \cite{ZGOZ} for more details).

More recently, the present authors \cite{ZGOZ} developed a microscopic theory
describing the QPS phenomenon and demonstrated that in sufficiently thin wires
QPS effects are well in the measurable range and may lead to a nonzero wire
resistivity even at $T=0$. Moreover, the existence of a new
superconductor-to-metal (insulator) phase transition as a function of the wire
thickness was pointed out in \cite{ZGOZ}.

In a present paper we extend our theory \cite{ZGOZ} in several important
aspects, in particular providing a more detailed discussion of the
QPS action in various limits and paying
attention on dissipative effects outside the QPS core. We also discuss a
possible explanation of the recently observed negative magnetoresistance 
\cite{Dynesnew} within the framework of our QPS scenario.

{\it The model.}
Our calculation is based on the effective action approach for a BCS
superconductor~\cite{sz}. The starting point is the partition function $Z$
expressed as an imaginary time path-integral over the electronic fields $\psi$
and the gauge fields $V, {\bf A}$, with Euclidean action
\begin{eqnarray}\nonumber
	S= \int d^{3}{\bf r}\int^{\beta}_{0}d\tau {\Big (}
	\bar{\psi}_{\sigma}
	[\partial_{\tau}-ieV+\xi({\bf \nabla}-ie{\bf A}/c)]
	\psi_{\sigma}  - \\ \nonumber
	-g\bar{\psi}_{\uparrow}\bar{\psi}_{\downarrow}
	\psi_{\downarrow}\psi_{\uparrow} + ieVn_{i}+
	[{\bf E}^{2}+{\bf B}^{2}]/8\pi	{\Big )} \; .
\end{eqnarray}
Here $\beta =1/T$, $\xi({\bf \nabla})\equiv -{\bf \nabla}^{2}/2m - \mu$, 
$en_{i}$ denotes the background charge density of the ions, and 
$\hbar=k_{B}=1$. A Hubbard-Stratonovich transformation introduces the
energy gap $\Delta$ as an order parameter and the electronic degrees of
freedom can be integrated out. What remains is an expression for the partition
function in terms of an effective action for $\Delta$, $V$ and ${\bf A}$, with
a saddle-point solution $|\Delta |=\Delta_{0}$ and $V={\bf A}=0$. We obtain
\begin{eqnarray}\nonumber
	S_{\rm eff}=\int d^{3}{\bf r}\int^{\beta}_{0}d\tau
	\left[\frac{|\Delta|^2}{g}+\frac{{\bf E}^2+{\bf B}^2}{8\pi} \right]-
		\mbox{Tr}\ln\hat{G}^{-1} \; ,\\ \nonumber
	\hat{G}^{-1}=\left(\partial_{\tau}+\frac{i}{2}
	\{{\bf \nabla},{\bf v}_{s}\}\right)\hat{1}+|\Delta |\hat{\sigma}_{1}+
	\\ \nonumber
	+\left(\xi({\bf \nabla})+\frac{m{\bf v}^{2}_{s}}{2}-ie\Phi\right)
	\hat{\sigma}_{3}\;,
\end{eqnarray}
where the superfluid velocity ${\bf v}_{s}=(1/2m)[{\bf \nabla}\varphi-
2e{\bf A}/c]$, the chemical potential for Cooper pairs $\Phi=V-
\dot{\varphi}/2e$, and $\Delta=|\Delta |e^{i\varphi}$ have been introduced.

{\it Effective action for QPS.}
The effective theory is constructed by expanding up to second order around the
saddle point in $\Phi$ and ${\bf v}_{s}$ to obtain the electronic
polarization terms \cite{ZGOZ,ogzb}.
A phase-slip event in imaginary time involves a suppression of the order
parameter in the phase slip core (i.e. inside the space-time domain
$x\leq x_{0}$, $\tau\leq \tau_{0} $), and a
winding of the superconducting phase around this core. The total QPS action
$S_{QPS}$ can be presented as a sum of a core part $S_{\rm core}$ around the
phase slip center for which the condensation energy and dissipation by normal
currents are important, and a
hydrodynamic part outside the core $S_{\rm out}$ which depends on the
hydrodynamics of the electromagnetic fields and
dissipation due to the presence of quasiparticles
above the superconducting gap.

In what follows we will consider sufficiently thin wires with the cross section
$S < \lambda_L^2$, where $\lambda_L$ is the London penetration length of a
bulk superconductor. Due to scattering on impurities and boundary imperfections
the electron mean free path $l$ in such wires is typically much shorter than
the coherence length of a clean sample $l \ll \xi_0=v_F/2\Delta$. Here
we restrict our attention to this physically important diffusive limit.
Assuming that outside the QPS core the
magnitude of the order parameter field is not suppressed
$|\Delta | =\Delta_{0}$ for $S_{\text{out}}$ we obtain \cite{ZGOZ,ogzb}
$$
S_{\rm out}=\!\int\!dxd\tau
{\Big (} \frac{C+C'}{2}V^{2}+ \frac{\tilde{C}}{2}\Phi^{2}+
\frac{1}{2Lc^{2}}A^{2}+ \frac{m^{2}{\bf v}^{2}_{s}}{2e^{2}\tilde{L}} {\Big )}
$$
\begin{eqnarray}
+\frac{S}{2\beta}\sum_{\omega < 1/\tau_0} \int dx \frac{\sigma
(\omega )}{|\omega |} |\partial_xV(\omega ,x)+ i\omega A(\omega
,x)/c|^2,
\label{Sout}
\end{eqnarray}
where the integration runs over $|x|>x_0, |\tau |>\tau_0$.
In general the kinetic inductance $\tilde L$ and the kinetic
capacitance $\tilde C$ in (\ref{Sout}) depend on frequency $\omega$ and the
wave vector $k$ \cite{ogzb}. In the limit of low $\omega$ and small $k$ 
we have $\tilde L =4\pi \lambda_L^2/S$ and $\tilde{C}=Se^{2}N_{0}n_{s}/n$,
where $n_s$ and $n$ are respectively the superconducting and the total electron
density. In (\ref{Sout}) we also introduced the
capacitance $C'=Se^{2}N_{0}n_{n}/n$ which we will drop from now on in the limit
$n_{s}\gg n_{n}\equiv n-n_s$ at low $T$.

The geometry and screening by dielectrics outside the wire are accounted
for by the capacitance per length $C$ and the inductance times length $L$
that replace the ${\bf E}^{2}+{\bf B}^{2}$-term.
[For thin wires transverse screening is irrelevant and we can retain only one
component of the vector potential]. The expressions
for $C$ and $L$ also depend on the relevant space and time scales as well
as on the wire geometry. In the ideal case of a cylindrical {\it uniform}
wire for $kr_0 \ll 1$ ($r_0$ is the wire radius) one has
$C=\epsilon_{r} [2\ln(/kr_{0})]^{-1}$ and $L=2\ln(/kr_{0}) /c^{2}$, $c$
is the speed of light and $\epsilon_{r}$ the dielectric constant of the
substrate.  In practice the details of the wire geometry can be very complicated
(e.g. the cross section $S$ is not constant along the wire, i.e. the wire 
is never uniform) and, on top of that, other (metallic) objects 
can be located in the vicinity of the wire.
The above effects lead to an
effective cutoff of the logarithmic dependence on $k$ at the scale $k \sim 1/d$
with $d$ depending on experimental details (e.g. $d$ can be a typical scale of
the wire inhomogeneity or the distance to the metallic groundplane). Here we
will stick to a simplified model and assume
$C=\epsilon_{r} [2\ln(d/r_{0})]^{-1}$ to be constant at all relevant distances.
As to $L$, its particular form turns out to be unimportant for thin wires with
$\sqrt{S} < \lambda_L$ in which case the kinetic inductance always dominates
$\tilde L \gg L$.
In addition to the above kinetic and electromagnetic effects the expression
(\ref{Sout}) accounts for dissipative currents outside the core. The
corresponding contribution is described by the last term in Eq. (\ref{Sout}).

As to the core contribution, it consists of two terms
\begin{eqnarray}
        S_{\rm core}=\frac{b}{2}N_{0}\Delta^{2}_{0}S\tau_{0}x_{0}+
	\frac{S}{\beta}\sum_{|\omega |>\tau^{-1}_{0}}\frac{x_{0}\sigma}
	{|\omega |}|E(\omega ,\frac{x_{0}}{2})|^{2} \; .
\label{Score}
\end{eqnarray}
The first part is the condensation energy that is lost inside the core and the
second part defines the energy of dissipative currents in the core
during a phase slip event. Here $\sigma$ is the normal state conductance of the
wire: we already made use of the fact that the typical QPS frequency is
sufficiently high \cite{ZGOZ} $1/\tau_0 \gtrsim \Delta_0$, therefore
dissipative currents inside the core are insensitive to superconductivity.
It is also important to emphasize that no
gradient terms for $\Delta$ (both in space and in time) should be added to
(\ref{Score}).  Such terms can be recovered only by expanding the effective
action in powers of $\omega$ and $k$. For fast processes (like QPS) this
expansion becomes obviously incorrect and it is necessary to carry out a
more careful treatment of polarization terms in the action. For the QPS event
with $1/\tau_0 > \Delta_0$ this treatment yields \cite{ogzb} $b \sim \ln
(1/(2\Delta_0\tau_0)+\xi^2/x_0^2)$, where $\xi = \sqrt{D/2\Delta_0}$ is the
coherence length of a dirty superconductor, $D=v_Fl/3$.

{\it Variational procedure}.
In order to evaluate the QPS action $S_{QPS}=S_{core}+S_{\text{out}}$ we will
use a variational approach which consists of several steps.

We first minimize the hydrodynamic contribution $S_{\text{out}}$
with respect to the potentials $V$ and $A$. As a result we arrive at the
saddle point conditions which link the potentials to the phase variable
outside the core.
Making use of the fact that for thin wires one has $\tilde L \gg L$,
$\tilde C \gg C$, we obtain in the Fourier representation 
\begin{eqnarray}
V_{\omega ,k}=\frac{i\omega\varphi_{\omega ,k}/2e}{1+ \sigma (\omega ,k)Sk^2/\tilde C|\omega |},
\label{1}\\
A_{\omega ,k}=-ikcL\varphi_{\omega ,k}/2e\tilde L.
\label{2}
\end{eqnarray}
With the aid of (\ref{1},\ref{2}) one can rewrite the action
$S_{\text{out}}$ only in terms of the phase variable $\varphi (\tau ,x)$.
Then minimizing this part of the action with respect to $\varphi$ and
keeping in mind the identity $\partial_x\partial_{\tau}\varphi
-\partial_{\tau}\partial_{x}\varphi =2\pi\delta (\tau ,x)$
(which follows from the fact that after a wind around the QPS center the
phase should change by $2\pi$) we find
\begin{equation}
S_{\text{out}}= \int_{|\omega |<1/\tau_0}\frac{d\omega }{2\pi }
\int_{|k|<1/x_0} \frac{dk}{2\pi }{\cal G}(\omega ,k).
\label{G}
\end{equation}
The general expression for the function ${\cal G}$ in (\ref{G}) is somewhat
tedious and is not presented here. In the following limits substantial
simplifications can be achieved:

i) $ \omega \gg \sigma (\omega ,k)Sk^2/C$.
The function (\ref{G}) has the form
\begin{equation}
{\cal G}(\omega ,k)=\frac{\pi^2/2e^2}{k^2/C+\tilde L\omega^2}.
\label{G1}
\end{equation}

ii) $ \sigma (\omega ,k)Sk^2/\tilde C \ll \omega \ll \sigma (\omega ,k)Sk^2/C$.
We find
\begin{equation}
{\cal G}(\omega ,k)=\frac{\pi^2\sigma (\omega ,k)S}{2e^2|\omega |}\frac1{1+
\tilde L\sigma (\omega ,k)S|\omega |}.
\label{G2}
\end{equation}

iii) $ \omega \ll \sigma (\omega ,k)Sk^2/\tilde C$. The function ${\cal G}$
again acquires the form (\ref{G1}) but with $C$ substituted by $\tilde C$.

As a last step of our variational procedure we minimize the total QPS
action $S_{\text{core}}+S_{\text{out}}$ with respect to the core parameters
$\tau_0$ and $x_0$.  Let us first neglect dissipation by formally putting
$\sigma =0$ in (\ref{Score}) and (\ref{G}). Then solving the equations
$$
\partial S_{QPS}/\partial \tau_0=0,\;\;
\partial S_{QPS}/\partial x_0=0
$$
and treating $b$ as a constant (i.e. neglecting its weak dependence on $\tau_0$
and $x_0$) we obtain
\begin{equation}
x_0=c_0\tau_0=\sqrt{\pi/4e^2\tilde LSbN_0\Delta_0^2},
\label{x01}
\end{equation}
where $c_0=1/\sqrt{\tilde LC}$ is the velocity of the Mooij-Sch\"on mode
\cite{ms} which determines the space-time asymmetry of the core.
The total action for a single QPS reads
\begin{equation}
S_{QPS}^{(0)}=\frac{\mu}2+\mu\ln\left(\frac{R}{2x_0}+\frac{R}{2c_0\tau_0}\right),
\label{QPS1}
\end{equation}
where $R^2 = X^2+c_0^2\beta^2$, $X$ is the wire length and 
$\mu =(\pi /4e^2)\sqrt{C/\tilde L}$.
The first term in (\ref{QPS1}) represents the core action $S_{\text{core}}$,
the second term defines $S_{\text{out}}$ (for simplicity we
chose the cutoff by integrating outside the ellipse 
$(x/x_0)^2+(\tau /\tau_0)^2>1$). 
Substituting $\tilde L = 4\pi\lambda_L^2/S= 1/2\pi e^2 N_0\Delta_0 DS$ into
(\ref{x01}) at $T \ll \Delta_0$ we find $x_0=c_0\tau_0= \pi \xi /\sqrt{b}$, 
i.e. the core size $x_0$ is of the order of the superconducting coherence length $\xi$,
and the QPS time $\tau_0 \sim \xi /c_0 \ll 1/\Delta_0$. This result justifies
the above conjecture that the typical QPS frequency is higher than $\Delta_0$
and demonstrates  why our core action $S_{\text{core}} \propto
x_0\tau_0$ is much smaller than that found within the TDGL analysis
\cite{Duan} which yields the QPS frequency of order $\Delta_0$.

Let us now include dissipation. 
At high frequencies dissipative currents flowing 
both inside and outside the core are important and should be taken into 
account even at $T=0$. 
The dissipative contribution from $S_{\text{out}}$ is obtained
from (\ref{G}), (\ref{G2}). After a simple integration one finds
\begin{equation}
S_{\text{out}}^{diss} \approx \sigma S/e^2x_0.
\label{Sdiss1}
\end{equation}
This expression is nothing but the Caldeira-Leggett dissipative action
of a normal conductor with the cross section $S$ and the length
$\sim x_0$. Similar expression 
defines the dissipative contribution from the core $S_{\text{core}}^{diss}$.

If $\sigma$ is small one can treat dissipative terms
perturbatively. This is sufficient as long as 
$\sigma S \lesssim e^2 \mu \xi$. It is easy to check that in 
the practically important Drude limit $\sigma =2e^2N_0D$ the above
condition would mean $\xi \gtrsim c_0/\Delta_0$. This condition is never
satisfied for realistic parameters. Therefore in this limit dissipation
cannot be treated perturbatively and our variational procedure should
be modified. 
Under certain simplifying assumptions one can find
\begin{equation}
S_{\text{core}}^{diss} \approx 
\frac{\sigma S}{e^2x_0}\left[ \frac 1{2r^6}+r^6\right] \ln \frac{c_0x_0}{Dr} ,  
\label{SE3}
\end{equation}
where $r=c_0\tau_0/x_0$. The strong dependence of (\ref{SE3}) on $r$ enforces the
minimum condition $r \approx 1$, i.e. the asymmetry of the core remains 
approximately the same as in the underdamped limit. Under this condition the
whole action $S_{QPS}^{(0)}+S_{QPS}^{diss}$ can be easily minimized with
respect to $x_0$ and we obtain \cite{ZGOZ} $x_0 \approx c_0\tau_0 \approx 
\xi \sqrt{a}$ and
\begin{eqnarray}
	S_{\text{core}}\approx a\mu, \;\;\; a \approx (c_0/\Delta_{0}\xi)^{2/3}.
\label{Score2}
\end{eqnarray}   

{\it Metal-Superconductor phase transition.}
The next step is to consider a gas of QPS's in a superconducting wire. We also
assume that an applied current $I$ (much smaller than the depairing current)
is flowing through the wire. Substituting the saddle point solution $\varphi=
\sum^{n}_{i}\tilde{\varphi}(x-x_{i},\tau-\tau_{i})$ into the action and
keeping track of the additional term $\int d\tau\int dx(I/2e)\partial_{x}
\varphi$~\cite{sz}, we find
\begin{equation}
        S_{n}=na\mu-\mu\sum\limits_{i\neq j}\nu_{i}\nu_{j}
	\ln \biggl{(}\frac{\rho_{ij}}{x_{0}}\biggr{)}+
	\frac{\Phi_{0}}{c}I\sum\limits_i\nu_{i}\tau_{i}\;.
\label{mul}
\end{equation}
The quantity
$\rho_{ij}=(c^{2}_{0}(\tau_{i}-\tau_{j})^2+(x_{i}-x_{j})^2)^{1/2}$
defines the distance between the i-th and j-th QPS in the $(x,\tau)$ plane,
$\nu_{i}=+1$ ($-1$) are the QPS (anti-QPS) ``charges'', and $\Phi_{0}=hc/2e$ is
the flux quantum. Only neutral QPS configurations with $\nu_{tot}= \sum_{i}^{n}
\nu_{i}=0$ (and hence $n$ even) contribute to the partition function \cite{ZGOZ}.

For $I=0$ Eq.~(\ref{mul}) defines the standard model of a 2D
gas of logarithmically interacting charges $\nu_{i}$. The effective (small)
fugacity $y$ of these charges is
\begin{equation}
        y=x_{0}\tau_{0}B\exp(-a\mu)\;,
\label{fug}
\end{equation}
where $B$ is the usual fluctuation determinant which we roughly
estimate as $B \sim a\mu /x_0\tau_0$. From
the Coulomb gas analogy, we conclude that a KTB phase transition~\cite{KTB}
for QPS's occurs in a superconducting wire at $\mu=\mu^{*}\equiv 2+4\pi y
\approx 2$: for $\mu<\mu^{*}$ the density of free QPS in the wire (and
therefore its resistance) always remains finite, whereas for $\mu >\mu^{*}$
QPS's and anti-QPS's (AQPS) are bound in pairs and the $linear$ resistance of
a superconducting wire is strongly suppressed and $T$-dependent. We
arrive at an {\it important conclusion}: at $T=0$ a 1D superconducting wire
has a vanishing linear resistance, provided
the electromagnetic interaction between phase slips is sufficiently strong,
i.e. $\mu>\mu^{*}$.

The above analysis is valid for sufficiently long wires. For typical
experimental parameters, however, $X<c_{0}\beta$ (or even $X\ll c_{0}\beta$),
and the finite wire size needs to be accounted for. Here we consider the 
physical situation with nonvanishing normal
conductivity of the wire $\sigma =\sigma_{qp}$ (and thus nonzero dissipation)
even far from the the QPS core. 
This situation can be realized in the presence of quasiparticles above the 
gap due to finite temperature and/or nonequilibrium effects. In this case 
our consideration should be modified as follows.

We first apply the 2D scaling ~\cite{KTB} $\partial_{l}y=(2-\mu)y$
and $\partial_{l}\mu=-4\pi^{2}\mu^{2}y^2$, 
where $\mu$ and $y$ depend on the scaling parameter $l$. 
Solving these equations up to $l=l_{X}=\ln(X/x_{0})$ we obtain the 
renormalized fugacity $\tilde{y}=y(l_{X})$. 

For larger scales $l>l_{X}$ only the time coordinate matters. At 
sufficiently low frequencies the inter-QPS interaction is determined
by the function (\ref{G2}) and
the problem reduces to a that of a 1D Coulomb gas with logarithmic
interaction. Therefore, (for $\tilde{y}\ll 1$) further scaling 
is defined by \cite{s,sz} $\partial_{l}\tilde y=(1-\gamma )\tilde y$ and 
$\partial_{l}\gamma =0$, where $\gamma = \pi S \sigma_{qp}/2e^2X$ is
the dimensionless ``quasiparticle'' conductance of the wire.
For $\gamma >1$ the fugacity scales down to zero, which again corresponds
to a superconducting phase, whereas for $\gamma <1$ it increases
indicating a resistive phase in complete analogy to a single Josephson
junction with ohmic dissipation. The phase transition point again depends on $S$,
but also on the wire length $X$ and the value $\sigma_{qp}$ (see below).

{\it Wire resistance at low T.}
At any nonzero $T$ the wire has a nonzero resistance $R(T,I)$ even in the
``ordered'' phase $\mu >\mu^{*}$ (or $\gamma  >1$). In order to evaluate
$R(T)$ in this phase for a long wire we proceed perturbatively and first
calculate the free energy correction $\delta F$ due to one bound QPS-AQPS
pair. The one QPS-AQPS pair contribution $\delta F$ to the free energy
of the wire is
\begin{equation}
	\delta F=\frac{Xy^{2}}{x_{0}\tau_{0}}\int^{\beta}_{\tau_{0}}
	\frac{d\tau}{\tau_{0}} \int^{X}_{x_{0}}\frac{dx}{x_{0}}
	e^{(\Phi_{0}I\tau/c)-2\mu\ln[\rho(\tau,x)/x_{0}]}\;,
\label{fren}
\end{equation}
where $\rho=(c^{2}_{0}\tau^{2}+x^{2})^{1/2}$. 
For nonzero $I$ the expression in Eq.~(\ref{fren})
is formally divergent for $\beta\rightarrow\infty$ and (after a proper
analytic continuation) acquires an imaginary part Im $\delta F$
This indicates a QPS-induced
instability of the superconducting state of the wire.
The corresponding decay rate $\Gamma =2\mbox{Im}\delta F$ defines the
total voltage drop $V$ across the wire (see \cite{ZGOZ} for more details).
For the wire resistance $R(T,I)=V/I$ this yields $R\propto T^{2\mu-3}$ and
$R\propto I^{2\mu-3}$ for $T \gg \Phi_{0}I$ and $T\ll\Phi_{0}I$ respectively.
For thick wires with $\mu>\mu^{*}$, we expect a strong temperature dependence
of the resistivity. For thinner wires the temperature dependence of the
resistivity becomes linear at the transition to the disordered phase in which
our analysis is not valid. At $T\ll\Phi_{0}I/c$ we expect a strongly nonlinear
I-V characteristic $V\sim I^{\nu}$ in thick wires, and a universal
$\nu(\mu^{*})=2$ in thin wires at the transition into the resistive state with
$V\sim I$, i.e. $\nu=1$. Note that in contrast to the KTB transition in 2D
superconducting films, the jump is from $\nu=2$ to 1, instead of $\nu=3$ to 1.

For a short wire $X<c_{0}/T$ we again proceed in two steps. A 2D scaling
analysis yields the ``global'' fugacity $\tilde{y}$. In analogy with the
resistively shunted Josephson junction~\cite{sz}, the voltage drop from the
imaginary part of the free energy reads
$$
        V=\frac{2\Phi_{0}\tilde{y}^{2}}{\Gamma(2\gamma )c\tilde{\tau}_{0}}
        \sinh\left(\frac{\Phi_{0}I}{2cT}\right)
        \vert\Gamma(\gamma +\frac{i\Phi_{0}I}{2\pi cT})\vert^{2}
        \left[\frac{2\pi\tilde{\tau}_{0}}
                {\beta}\right]^{2\gamma -1}\;,
$$
giving $R\propto T^{2\gamma -2}$ and $R\propto I^{2\gamma -2}$
respectively at high and low $T$. Here $\tilde \tau_0$ is
defined from the high frequency cutoff in (\ref{G2}): 
$\tilde \tau_0 \sim XC/e^2\gamma$. The above result is valid 
for $\gamma >1$ and
also for smaller $\gamma$ at not very small $T$~\cite{sz}. At $T\to 0$ in
the metallic phase the resistance becomes~\cite{sz}
\begin{equation}
        R=S\sigma_{qp}/X \;,
\label{Rshort0}
\end{equation}
i.e. $R$ is just equal to the quasiparticle resistance of the wire whereas 
the superconducting channel is blocked due to quantum fluctuations.

{\it Discussion}.
Let us compare our predictions with experimental results 
\cite{Gio,Dynes,Dynesnew}. Taking $\sqrt{S} \sim$ 10 nm and $\epsilon_{r}$ = 1,
for typical system parameters $k^{-1}_{F} \sim 0.2$ nm $<l\sim 1\div 10$ nm 
$\lesssim \xi
\sim 10$ nm $<\xi_{0}\sim \lambda_{L}\sim 100$ nm
 we obtain the velocity $c_{0}/c=c_{MS}/c\approx (\sqrt{S}/10\lambda_{L})$,
$\mu \approx 30(\sqrt{S}/\lambda_{L})$ and $a \sim 5 \div 10$.
This estimate yields the core action 
$S_{\text{core}}\simeq a\mu \lesssim 10$ in agreement
with \cite{Gio}. 

For the quoted parameters, we predict the superconductor to metal transition
at a wire thickness $\sqrt{S}\simeq\lambda_{L}/15 \lesssim 10$ nm. This
prediction also agrees with the results of Giordano, who finds that wires with
$r_0=\sqrt{S/\pi} \approx 8$ nm have a resistivity that 
saturates at a measurable level
at low $T$, whereas the resistivity of thicker wires \cite{Gio}
$r_0 \gtrsim 13$ nm always decreases with $T$. 

Another remarkable feature is that the classical-to-quantum
crossover temperature $T^*$ was found to be quite close to $T_C$ 
for sufficiently thin wires \cite{Gio}. Comparing our quantum action
$2S_{\text{core}}$ with the classical exponent \cite{LAMH} we immediately
arrive at a simple estimate for 
$T^{*}\approx \Delta_0^{2/3}c_0^{1/3}/\xi^{1/3}$. For the above
parameters it yields $T^{*}\sim 10\Delta(T^{*})$,
i.e. for thin wires one indeed expects this crossover 
to happen quite close to $T_C$.

Independent measurements of $R(T)$ for superconducting wires have
been carried out in Refs. \cite{Dynes,Dynesnew}, where systematic
deviations from classical predictions \cite{LAMH} for thin
wires have been also reported. Although the overall trend
\cite{Dynes,Dynesnew} is similar to that observed in \cite{Gio} 
the shape of some experimental curves look quite different from \cite{Gio}.
It was argued \cite{Duan} that these quantitative differences are
due to granularity of the wires used in the experiments \cite{Gio}.
However, the variations of $S$ were reported to be moderate
in \cite{Gio}. If so, this experimental feature can only cause
a somewhat nonuniform distribution of QPSs along the wire because the
QPS fugacity increases with decreasing $S$. With trivial
modifications our theory can be applied to this situation as well.

In order to proceed with our comparison let us recall that the wires 
in \cite{Dynes,Dynesnew} were quite short $X \sim 1\div 2 \mu$m 
whereas the wires investigated in \cite{Gio} were typically
by one or even two orders of magnitude longer. At not very low
$T$ one can put $\sigma_{qp} \sim \sigma$ and estimate the
parameter $\tilde \tau_0^{-1}$ to be of order 1 K or even bigger
for the samples \cite{Dynes,Dynesnew}. At the same time for
the samples \cite{Gio} $\tilde \tau_0^{-1}$ is typically below 10 mK.
Thus the difference between the results \cite{Gio} and 
\cite{Dynes,Dynesnew} can be attributed to the different behavior
of the function ${\cal G}$ (\ref{G}) for different frequencies:
the form (\ref{G1}) should be applied in the case \cite{Gio}
whereas the samples \cite{Dynes,Dynesnew} in the interesting
temperature range should be rather described by means of (\ref{G2}). 
E.g. it appears that the data of fig. 1a of \cite{Dynesnew}
can be (at least qualitatively) described by the dependence
\begin{equation}
R(T) \propto \tilde y^2(T\tilde \tau_0)^{2\gamma-2}
\label{R}
\end{equation}
both for thicker $\gamma >1$ and thinner $\gamma <1$ wires. The resistance
of the latter -- according to our analysis -- should increase with
decreasing $T$. This is just what has been found in \cite{Dynesnew}.
The crossover from this behavior to that with decreasing $R(T)$ for
thicker wires can be interpreted as an indication  of the phase
transition at $\gamma =1$.

Another interesting feature to be discussed is the negative magnetoresistance
of the wires observed in \cite{Dynesnew}. At the first sight this feature
contradicts to our QPS scenario: in a (sufficiently strong)
magnetic field $H$ the gap $\Delta_0$ is partially suppressed and the barrier
for QPS should decrease. Hence, the fugacity $y$ and the wire resistance $R$
-- in contrast to \cite{Dynesnew} -- should increase with $H$. 
However, if one includes dissipative effects outside the core into account
the picture can change drastically. Indeed $\sigma_{qp}=\sigma n_n/n$
strongly depends on the relation between $T$ and
$\Delta_0 (T,H)$. At sufficiently low $T$ a decrease of $\Delta$ may
lead to an {\it exponential} ($\sigma_{qp} \propto \exp (-\Delta_0/T)$) 
increase of the number of quasiparticles and -- therefore -- dissipation.
Thus we have two effects: the field $H$ increases the QPS fugacity $y$
but also increases dissipation $\gamma$ which suppresses quantum fluctuations.
It is quite obvious (e.g. from Eq. (\ref{R})) that the second effect 
may dominate in a certain parameter region and the
resistance will {\it decrease} with increasing $H$.
At very large $H$ the gap $\Delta_0$ will be suppressed
and the resistance will increase back. 
This reentrant behavior was observed in \cite{Dynesnew}.

{\it Note added.} A clear experimental evidence for a dissipative 
phase transition controlled by the normal state wire resistance $R$ 
was obtained very recently in ref. \cite{BT}. In these experiments several 
uniform wires with thicknesses in the 10 nm range were studied.
The wires with $R <R_q=\pi/2e^2 \simeq 6.5$ K$\Omega$ were found
to go superconducting at low $T$ while the resistance of the wires with $R >R_q$
showed no tendency to decrease with temperature even well below $T_C$.
It appears, therefore, that the results \cite{BT} can be
interpreted in terms of the dissipative phase transition at 
$\gamma =R_q/R=1$ predicted in our paper. We are grateful to A. Bezryadin
and M. Tinkham for communicating their results to us prior to publication.

\end{multicols}
\end{document}